\documentclass[useAMS,usenatbib]{mn2e}

\usepackage{graphicx}
\usepackage{txfonts}
\usepackage{natbib}
\newcommand\aj{AJ}
\newcommand\apj{ApJ}
\newcommand\apjs{ApJS}
\newcommand\apss{Ap\&SS}
\newcommand\aap{A\&A}
\newcommand\mnras{MNRAS}
\newcommand\apjl{ApJ}
\newcommand\pasp{PASP}
\newcommand\pasj{PASJ}

\title[]{The Binary Populations of Eight Globular Clusters in the Outer Halo of the Milky Way}
\author[A.\,P.\,Milone et al.]
{A.\,P.\,Milone$^{1}$, 
A.\,F.\,Marino$^{1}$, 
L.\,R.\,Bedin$^{2}$, 
A.\,Dotter$^{1}$, 
H.\,Jerjen$^{1}$,
D.\,Kim$^{1}$,
\newauthor
D.\,Nardiello$^{2,3}$,
G.\,Piotto$^{2,3}$, 
J.\,Cong$^{1}$
\\ 
$^{1}$Research School of Astronomy \& Astrophysics, Australian National University, Mt Stromlo Observatory, via Cotter Rd, Weston, ACT 2611, Australia \\
$^{2}$Istituto Nazionale di Astrofisica - Osservatorio Astronomico di Padova, Vicolo dell'Osservatorio 5, Padova, IT-35122\\
$^{3}$Dipartimento di Fisica e Astronomia ``Galileo Galilei'', Univ. di Padova, Vicolo dell'Osservatorio 3, Padova, IT-35122\\
 }

\begin{document}
\date{Accepted 2015 October 26.  Received 2015 October 1; in original form 2015 June 3}

\pagerange{\pageref{firstpage}--\pageref{lastpage}} \pubyear{2013}

\maketitle
\label{firstpage}
 
\begin{abstract}  
We analyse color-magnitude diagrams of eight Globular Clusters (GCs) in the outer Galactic
Halo. Images were taken with the Wide Field Channel of the Advanced Camera for Survey and the 
Ultraviolet and Visual Channel of the Wide Field Camera 3 on board of the {\it Hubble Space Telescope}. 
We have determined the fraction of binary stars along the main sequence and combined results 
with those of a recent paper where some of us have performed a similar analysis on 59 Galactic GCs.
 In total, binaries have been now studied homogeneously in 67 GCs.

We studied the radial and luminosity distributions of the binary systems, the distribution of their mass-ratios 
and investigated univariate relations with several parameters of the host GCs.
We confirm the anti-correlation between the binary fraction and the luminosity of the host cluster,
and find that low-luminosity clusters can host a large population in excess of $\sim$40\% in the cluster core.
However, our results do not support a significant correlation with the cluster age as suggested in the literature.

In most GCs, binaries are more centrally concentrated than single stars. If the fraction of binaries 
is normalised to the core binary fraction the radial density profiles follow a common trend. It has a maximum 
in the center and declines by a factor of two at a distance of about two core radii from the cluster center. 
After dropping to its minimum at a radial distance of $\sim$5 core radii it stays approximately constant at
larger radii. We also find that the mass-ratio and the distribution of binaries as a function of
the  mass of the primary star is almost flat.

\end{abstract}


\section{Introduction}\label{sec:intro}
An appropriate analysis of the color-magnitude diagram (CMD) of globular clusters (GCs) can provide an efficient tool to study their population of binaries stars (e.g.\,Romani \& Weinberg\, 1991; Bolte\, 1992; Rubenstein \& Baylin\, 1997; Bellazzini et al.\, 2002; Clark\, Sandquist \& Bolte\, 2004; Richer et al.\, 2004; Zhao \& Baylin\, 2005; Milone et al\, 2009; Ji \& Bregman\,2013, 2015).  

Such an approach can provide statistically robust results, because it allows the study of thousands of stars from a single CMD, and is sensitive to binary systems with any inclination and orbital period.
An accurate study of binaries requires high-precision photometry of stars in the crowded field of a GC, correction of differential reddening, and accurate analysis of photometric errors and field-stars contamination.
While previous studies included a small number of GCs, in the last years, accurate photometry from high-resolution {\it Hubble Space Telescope} ({\it HST}) images allowed the  systematic investigations of large sample of clusters (Sollima et al.\,2007; Milone et al.\,2012, hereafter MPB+12).

 MPB+12 investigated main-sequence (MS) binaries from homogeneous photometry of 59 Galactic GCs observed with the Wide Field Channel of the Advanced Camera for Surveys (WFC/ACS) of {\it HST} as part of the Globular Cluster Treasury project (GO\,10775, PI.\,A.\,Sarajedini, see Sarajedini et al.\, 2007).
The authors estimated the fraction of binaries, determined the mass-ratio and the radial distribution, and investigated relations between the fraction of binaries and the main parameters of the host GCs. Most GCs contain a fraction of binaries smaller than the fraction of binaries in the Galactic field. Binaries are more centrally concentrated than single stars, with the fraction of binaries generally dropping down by a factor of two from the center to about two core-radii. There is a significant anticorrelation between the fraction of binaries and the cluster mass.

All the GCs studied by Sollima et al.\,(2007) and MPB+12 and included in the Globular Cluster Treasury project have Galactocentric distances smaller than $R_{\rm GC} \sim$21\,kpc. The population of binaries in outer-halo GCs is almost unexplored.
 Archive data taken with WFC/ACS and with the Ultraviolet and Visual Channel of the Wide Field Camera 3 (UVIS/WFC3) are available for eight outer-halo GCs, namely Pyxis, Ruprecht\,106, IC\,4499, NGC\,6426, NGC\,7006, Palomar\,15, AM\,4, and Palomar\,13. These images have been collected through the same F606W and F814W filters as used in GO\,10775 \footnote{ The photometry of MS stars in the F606W and F814W of WFC/ACS and the UVIS/WFC3 filters is very similar. Isochrones from Dotter et al.\,(2008) with age and metallicity of AM\,4 ([Fe/H]=$-$1.3, Harris\,1996, 2010, age=13\,Gyr) show that the maximum difference between the luminosity of MS stars brighter than  $m_{\rm F814W, TO}+3.75$, where $m_{\rm F814W, TO}$ is the magnitude of the main-sequence turn off, is $\Delta m_{\rm F606W}=$0.018 and $\Delta m_{\rm F814W}=$0.011 mag. The maximum color difference is $\Delta m_{\rm F606W}-m_{\rm F814W}=0.008 mag$. When we use isochrones with the metallicity as Pal\,13 ([Fe/H]=-1.88, Harris\,1996, 2010) we obtain identical results: $\Delta m_{\rm F606W}=$0.016, $\Delta m_{\rm F814W}=$0.010 mag, and $\Delta m_{\rm F606W}-m_{\rm F814W}=0.006 mag$. Such small differences are negligible for our purposes.}. 

 In this paper, we exploit this dataset to extend the study of binaries along the MS by MPB+12 to outer-halo GCs.

\section{Data and data analysis}
 To determine the fraction of binaries in AM\,4 and Pal\,13 we have used photometry and astrometry from images collected with the UVIS/WFC3 on board of {\it HST} (GO\,11680, PI:\,G.\,H.\,Smith, see Hamren et al.\,2013). This dataset consists of 4$\times$615s in F606W and 4$\times$620s in F814W for AM\,4 and 4$\times$610s in F606W and 4$\times$615s in F814W for Pal\,13.
   For the other six outer-halo GCs we have used the photometric catalogs from Dotter et al.\,(2011) obtained by using ACS/WFC images from GO\,11586 (PI:\,A.\,Dotter, see Dotter et al.\,2011).

UVIS/WFC3 images have been corrected for the effect of poor charge transfer efficiency following Anderson \& Bedin\,(2010)  and by using the software written and provided by these authors. Briefly, they have developed a model that reproduces the trails observed in a large sample of analysed dark exposures. Their software converts the observed into an estimate of the original pixel values by inverting the derived model.

The software that we have adopted to obtain photometry and astrometry of stars in UVIS/WFC3 images  has been provided to us by Jay Anderson, is based on the recipe by Anderson et al.\,(2008) and has been adapted to UVIS/WFC3 by Jay Anderson.
  We fitted to each star in each exposure a grid of PSFs interpolated specifically for that star, from a 9$\times$5  PSF grid in each chip of the UVIS/WFC3.
This grid model consists of a library PSF by Anderson et al.\,(in preparation) plus  a spatially-constant perturbation for each exposure which accounts for any focus variation due to the `breathing' of {\it HST}. The PSF fitting of each isolated star uses the central 5$\times$5 pixel and is performed after subtracting off its neighbours as explained in Anderson \& King\,(2006). In the case of stars with significant contamination from close neighbours we performed the PSF fitting on the centermost pixels. The sky level is measured from an annulus with radii 4 and 8 pixels for the brighter stars and 3 and 7 pixels for the fainter stars.  
 As described in detail by Anderson et al.\,(2008), we have inferred positions and luminosities of bright and faint stars by using two different methods. Bright stars have been measured independently in each image and results combined later, while astrometry and photometry of each very faint star has been determined by simultaneously fitting all the pixels in all the exposures.

Stellar positions have been corrected for geometrical distortion using the solution by Bellini, Anderson \& Bedin\,(2011), and photometry has been calibrated to the Vega-mag system as in Bedin et al.\,(2005) by using the zero points provided by the STScI web page for WFC3/UVIS\footnote{http://www.stsci.edu/hst/acs/analysis/zeropoints/zpt.py}. 

We exploit three indexes provided by the adopted software and described in detail by Anderson et al.\,(2008) as diagnostics of photometric quality:
  the fraction of flux in the stellar aperture from its neighbours ($o$), the fractional residuals in the fit of the PSF to the pixel ($q$), and the rms of the independent measurements of the average position in the F606W and F814W images.
 We limited our study to relatively-isolated stars with small astrometric uncertainties and well fitted by the PSF, in close analogy to MPB+12. These stars have been selected as in MPB+12 (see their Sect.~2.1), by rejecting any source that, in comparison to stars with similar luminosity, exhibit anomalously large values of the $q$ or $o$ parameters, or large values of the position rms.

Finally, the CMDs have been corrected for differential reddening and for spatial variation of the photometric zero point due to small PSF inaccuracies by using the method described in MPB+12 (see their Sect.~3).
 Briefly, we have first rotated the CMD of each cluster in a way that the new abscissa is parallel to the reddening vector.  Then, we have derived a fiducial line of the MS and the SGB of each cluster, by putting a spline through the median value of the quantities in the abscissa (pseudo-color) and the ordinata (pseudo-magnitude) in the rotated reference frame derived in successive short bins of pseudo-magnitude, and we iterated this step with a sigma clipping; 
 We have then derived for each star the pseudo-color residuals of a sample of 50 relatively-bright and well-measured MS and SGB stars with respect to the fiducial line. To do this we have excluded the target star from the calculation of its own differential reddening. We assumed as differential reddening of each star, the median value of such residuals measured along the reddening line.

\subsection{Artificial Stars} 
\label{subsec:ASs}
As discussed in MPB+12 and in the next subsection, artificial stars (ASs) are crucial to determine the fraction of binaries in any GC.
 AS-tests have been performed as in Anderson et al.\,(2008).
Briefly, we have first generated a list including the position in the reference frame and the magnitudes of  500,000 stars. ASs have been placed along the MS fiducial line in the CMD of each cluster  with a flat luminosity  function in the F814W band covering the magnitude range from the saturation level down
 to an instrumental F814W magnitude\footnote{instrumental magnitudes are defined as $-$2.5 log$_{10}$(flux$_{\rm DN}$)} of $-$4.5. 
The projected distribution of stars is flat within the cluster core and declines as $r^{-1}$ outside the core.

The software described by Anderson et al.\,(2008) generates for each star in the input list and for each image, a star with the appropriate coordinates and flux. Then it measures its magnitude and position by using the same procedure as for real stars and provides the same diagnostics of the photometric quality as for real stars.  We considered an AS as recovered when the measured luminosity in each band differs from the input value by not more than 0.75 mag, and the position by less than 0.5 pix in each coordinate. We used for ASs the same procedure described above for real stars to select a sub-sample of relatively-isolated stars with small astrometric errors, and well fitted by the PSF. 
 ASs have been first used to derive the completeness level of our sample by following the recipe in MPB+12 (see their Sect.~2.2 for details). Briefly, we have determined a $5\time8$ grid in the magnitude-radius plane. Specifically we have subdivided the field of view into five circular regions and within each of them, we have analysed AS results in 8 magnitude bins over the interval from $-13.75<m_{\rm F814W}<-4.5$. We have determined, for each grid point, the completeness as the fraction of recovered to added stars and interpolated these grid points to estimate the completeness value associated to each star.
 Moreover, a subsample of 100,000 ASs have been used to estimate the fraction of chance-superposition binaries as we will descrive in the next Section.

\subsection{The fraction of binaries}
In this section we describe the method adopted to determine the fraction of binaries with mass ratio, q$>$0.5. Indeed, binaries with smaller mass ratios are very close to the MS fiducial line and, due to photometric errors, are indistinguishable from single MS stars. 

\begin{centering}
\begin{figure*}
 \includegraphics[width=15cm]{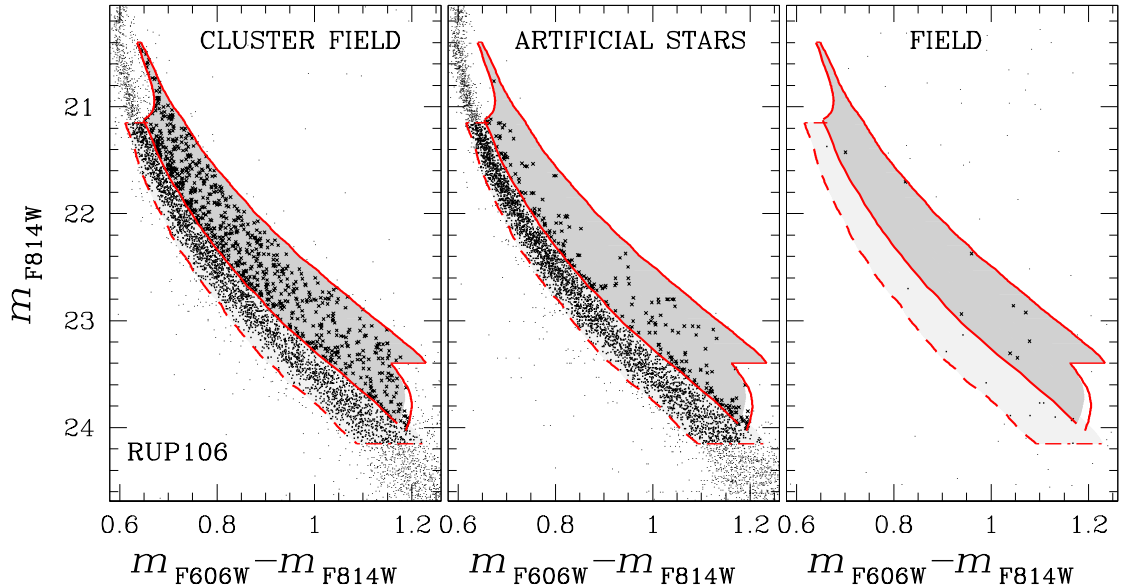}
 \caption{This figure illustrate the procedure to determine the fraction of binaries with q$>$0.5 for Rup\,106. The CMD of stars in the cluster field, the CMD obtained from ASs, and the CMD of field stars are plotted in the left, middle, and right panel, respectively.  The region A of the CMD includes both light- and dark-gray areas, while region B is the subregion of A coloured dark gray. Stars in the region A are marked either with black dots or with black crosses. In particular, black crosses indicate stars in the region B.}
 \label{fig:setup}
\end{figure*}
\end{centering}

To determine the fraction of binaries with mass ratio q$>$0.5 we used the
method described by MPB+12, that we summarise here for convenience.
Briefly, we determined the magnitude of the MS turnoff (TO) in the F814W band ($m_{\rm F814W, TO}$) and defined two regions in the CMD: a region ``A''
populated by single stars and those binary systems with a primary component with
$m_{\rm F814W, TO}+0.75<m_{\rm F814W}<m_{\rm F814W, TO}+3.75$ (the shadowed area in each panel of Fig.~\ref{fig:setup}), and a region ``B'' which is the portion of A containing the binaries with q$>$0.5 (the darker area in the
panels of Fig.~\ref{fig:setup}). The reddest line is the locus of the
equal-mass binaries  shifted to the red by 4-$\sigma$ (where $\sigma$ is the
error estimated as in MPB+12). The dashed line is the MS
fiducial moved by  4 $\sigma$ to the blue. The locus of the CMD of
binaries with a given mass ratio has been determined by using the
mass-luminosity relation provided by the best-fitting isochrones from Dotter et al.\,(2008).

The fraction of binaries with q$>$0.5 is calculated as in Eq.~1 in
MPB+12:
\begin{equation}
\label{eq:1}
f_{\rm bin}^{\rm q>0.5}=\frac {N_{\rm REAL}^{\rm B}-N_{\rm FIELD}^{\rm B}}
      {N_{\rm REAL}^{\rm A}-N_{\rm FIELD}^{\rm A}} - \frac {N_{\rm
          ART}^{\rm B}}{N_{\rm ART}^{\rm A}}, 
\end{equation}

where $N_{\rm REAL}^{\rm A,(B)}$ is the number of cluster stars (corrected for completeness) observed in region A (B) of the CMD; $N_{\rm ART}^{\rm A,(B)}$ and $N_{\rm FIELD}^{\rm A,(B)}$ are the corresponding numbers of artificial stars, and field stars normalised to the area of the cluster field.  

 In order to derive the CMD of artificial stars used to estimate $N_{\rm ART}^{\rm A,(B)}$ we have used ASs. Specifically, we have selected for each observed star a set of ASs from the catalog derived in Sect.~\ref{subsec:ASs} with a measured F814W magnitude within 0.2 mag and a radial distance within 200 pixels for a total 100,000 ASs. This procedure ensures that the sample of artificial stars used to infer the fraction of chance-superposition binaries, that is the ratio $\frac {N_{\rm ART}^{\rm B}}{N_{\rm ART}^{\rm A}}$, has almost the same spatial and luminosity distribution as the observed ones. The CMD derived from ASs is shown in the middle panel of Fig.~\ref{fig:setup} for Rup\,106 where for clearness, we have selected a subsample of  stars that is equal in number to the number of real stars plotted in the right panel of the same figure.  
To estimate the number of field stars we used a synthetic $m_{\rm F814W}$ versus $m_{\rm F606W}-m_{\rm F814W}$ CMD containing the same number of field stars expected in the field of view of each GC.  To do this, we used the program  Trilegal 1.6\footnote{http://stev.oapd.inaf.it/cgi-bin/trilegal} which predicts star counts, and colours and magnitudes for stars in any Galactic field on the basis of the Galactic model by Girardi et al.\,(2005).
We repeated the same procedure for the fraction of binaries with q$>$0.6 and q$>$0.7.

As done in MPB+12, we have analysed annuli with different radial distances from the cluster center.  When we study the binaries in a given annulus, the ASs used to derive the fraction of binaries as in Eq.~\ref{eq:1} have been extracted from the same annulus in such a way that the photometric and astrometric measurements of both real and artificial stars are similarly affected by crowding.
 We have measured both the total fraction of binaries and the fraction of binaries  with q$>$0.5 in the core ($r_{\rm C}$ sample), in the region between the core and the half-mass radius ($r_{\rm C-HM}$ sample), and outside the half-mass radius ($r_{\rm oHM}$ sample). 
 The values for the core radii are from the Harris\,(1996, updated as in 2010) catalog while the values for the half-mass radii are from the 2003 edition. The clusters centers have been determined by Ryan Goldsbury (private communication) as in Goldsbury et al.\,(2010) and will be published in a separate paper. In the case of NGC\,7006, which is the most-massive cluster of our sample, we have excluded from the analysis stars within a minimum cluster radius ($R_{\rm MIN}$=0.8 arcmin) where crowding prevents us from distinguishing binaries with q$>0.5$.

In principle, the total fraction of binaries, $f_{\rm bin}^{\rm TOT}$ can be inferred from the fraction of binaries by adopting a mass-ratio distribution (Sollima et al.\,2007). In this paper, we will follow the approach by Sollima and collaborators and estimate $f_{\rm bin}^{\rm TOT}$ by extrapolating the values obtained for $f_{\rm bin}^{\rm q>0.5}$. 
 MPB+12 compared the binary fraction determined for 59 GCs in small intervals of mass ratio and concluded that the analysed GCs have, on average a flat mass-ratio distribution for q$>$0.5. Unfortunately, the mass-ratio distribution of binaries in GCs is poorly constrained for q$<$0.5.
 In this paper we will extrapolate the result by MPB+12 to lower mass ratio and assume that the total fraction of binaries is two times the fraction of binaries with q$>$0.5. 
 For completeness, we recall the conclusion by MPB+12 that, if we assume a Fisher et al.\,(2005) mass-ratio distribution the total binary fraction is $f_{\rm bin}^{\rm TOT}=1.504 f_{\rm bin}^{\rm q>0.5}$, while if we assume that binary pairs formed by random association of stars with different masses according to a  Kroupa\,(2002) initial mass function, we obtain $f_{\rm bin}^{\rm TOT}=4.167 f_{\rm bin}^{\rm q>0.5}$.

\subsection{Globular-cluster parameters}\label{subs:parameters}
In Sect.~\ref{sec:relations} we will correlate the binary fraction with a number of physical and morphological parameters of the host GCs.
For that purpose we extracted the following parameters from the 2010 edition of the Harris\,(1996) catalog:
  absolute magnitude ($M_{\rm V}$), ellipticity ($e$), King\,(1962)-model central concentration ($c$), central velocity dispersion ($\sigma_{\rm V}$), logarithm of central stellar density ($\rho_{0}$), metallicity ([Fe/H]), core relaxation half time ($\tau_{\rm c}$), and half-mass relaxation half time ($\tau_{\rm hm}$), central surface brightness ($\mu_{\rm V}$), distance from the Galactic center ($R_{\rm GC}$), specific frequency of RR\,Lyrae ($S_{\rm RR~Lyrae}$). We used two sets of values for cluster ages. These include ages determined by the group of VandenBerg and collaborators (VandenBerg et al.\,2013 and Leaman et al.\,2013)  and ages based on Dartmouth isochrones, estimated by Dotter et al.\,(2010, 2011) and Hamren et al.\,(2013). In the case of AM\,4 we derived an age of 13.0$\pm$1.5 Gyr by adopting the same method and the same isochrones used by Dotter et al.\,(2010) and assuming  the same values of metallicity, reddening, and distance modulus as in Hamren et al.\,(2013).  In addition, we have used the values of the R-parameter ($R$) provided by Salaris et al.\,(2004), which are sensitive to the helium content of GCs. 

To study the relation between binaries and the horizontal branch (HB), we have used several  quantities as indicator of the HB morphology. The HB-ratio, (HBR), from the 2003 version of the Harris\,1996 catalog, 
 the $m_{\rm F606W}-m_{\rm F814W}$ color distance from the RGB and the reddest part of the HB ($L1$), and the $m_{\rm F606W}-m_{\rm F814W}$ color extension of the HB ($L2$) from Milone et al.\,(2014).

\section{Results}
 The fraction of binaries with q$>0.5$, q$>0.6$ and q$>0.7$ and the total fraction of binaries extrapolated by assuming a flat mass-ratio distribution is listed in Table~\ref{tab:risultati} for the ${\it r}_{\rm  C}$, ${\it r}_{\rm C-HM}$, and ${\it r}_{\rm oHM}$ sample and in  the whole field of view.
 These values will be used in Sect.~\ref{sec:relations} to investigate univariate relations between the fraction of binaries and a number of structural and morphological parameters of the host GC. In the following subsection we will combine results from this paper and from MPB+12 to analyse the radial distribution, the mass-ratio distribution and the distribution of the binary systems as a function of the luminosity and the mass of the primary component.
\begin{table*}
\label{tab:risultati}
\centering
\caption{Fraction of binaries with mass ratio q$>0.5$, q$>0.6$ and
  q$>0.7$, and total fraction of binaries measured in different regions.
}
\begin{tabular}{cccccc}
\hline
\hline
\noalign{\smallskip}
\multicolumn{1}{c}{ID} &  
\multicolumn{1}{c}{REGION} &  
\multicolumn{1}{c}{${\it f}_{\rm bin}^{\rm q>0.5}$} & 
\multicolumn{1}{c}{${\it f}_{\rm bin}^{\rm q>0.6}$} &    
\multicolumn{1}{c}{${\it f}_{\rm bin}^{\rm q>0.7}$} &     
\multicolumn{1}{c}{${\it f}_{\rm bin}^{\rm TOT}$} \\ 
\\
\hline
\noalign{\smallskip}
 AM\,4     & ${\it r}_{\rm C}$ sample     & 0.221$\pm$0.106  &  0.255$\pm$0.095 & 0.181$\pm$0.069 &  0.452$\pm$0.212 \\
           & ${\it r}_{\rm C-HM}$ sample     & ---  &  --- & --- & --- \\
           & ${\it r}_{\rm oHM}$ sample   & 0.202$\pm$0.058  &  0.153$\pm$0.044 &  0.145$\pm$0.037 &  0.404$\pm$0.116 \\
${\it R}_{\rm min}$=0.00 & UVIS field                  & 0.201$\pm$0.051  &  0.177$\pm$0.040 &  0.152$\pm$0.032 &  0.402$\pm$0.102 \\
\hline
 IC\,4499  & ${\it r}_{\rm C}$ sample    & 0.063$\pm$0.007  &  0.053$\pm$0.005 & 0.044$\pm$0.004 &  0.126$\pm$0.014 \\
           & ${\it r}_{\rm C-HM}$ sample  & 0.050$\pm$0.005  &  0.039$\pm$0.003 & 0.030$\pm$0.003 &  0.100$\pm$0.010 \\
           & ${\it r}_{\rm oHM}$ sample   & 0.042$\pm$0.009  &  0.035$\pm$0.007 &  0.027$\pm$0.005 &  0.084$\pm$0.018 \\
${\it R}_{\rm min}$=0.00 & WFC field                  & 0.052$\pm$0.004  &  0.042$\pm$0.003 &  0.034$\pm$0.002 &  0.104$\pm$0.008 \\
\hline
 NGC\,6426 & ${\it r}_{\rm C}$ sample    & 0.109$\pm$0.023  &  0.114$\pm$0.018 & 0.077$\pm$0.014 &  0.218$\pm$0.046 \\
           & ${\it r}_{\rm C-HM}$ sample  & 0.090$\pm$0.008 &  0.085$\pm$0.006 &  0.067$\pm$0.004 &  0.180$\pm$0.016 \\
           & ${\it r}_{\rm oHM}$ sample   & 0.070$\pm$0.007 &  0.054$\pm$0.005 &  0.040$\pm$0.003 &  0.140$\pm$0.014 \\
${\it R}_{\rm min}$=0.00 & WFC field                  & 0.080$\pm$0.005  &  0.071$\pm$0.004 &  0.053$\pm$0.003 &  0.160$\pm$0.010 \\
\hline
 NGC\,7006 & ${\it r}_{\rm C}$ sample     & ---  &  --- & --- & --- \\
           & ${\it r}_{\rm C-HM}$ sample   & ---  &  --- & --- & --- \\
           & ${\it r}_{\rm oHM}$ sample   & 0.029$\pm$0.009  &  0.021$\pm$0.006 &  0.017$\pm$0.004 &  0.058$\pm$0.012 \\
${\it R}_{\rm min}$=0.80 & WFC field       & 0.029$\pm$0.009  &  0.021$\pm$0.006 &  0.017$\pm$0.004 &  0.058$\pm$0.012 \\
\hline
 PAL\,13   & ${\it r}_{\rm C}$ sample    & 0.383$\pm$0.094  &  0.345$\pm$0.077 & 0.267$\pm$0.062 &  0.766$\pm$0.188 \\
           & ${\it r}_{\rm C-HM}$ sample    & ---  &  --- & --- & --- \\
           & ${\it r}_{\rm oHM}$ sample   & 0.302$\pm$0.044  &  0.247$\pm$0.035 &  0.179$\pm$0.028 &  0.604$\pm$0.088 \\
${\it R}_{\rm min}$=0.00 & UVIS field                  & 0.324$\pm$0.042  &  0.271$\pm$0.034 &  0.201$\pm$0.027 &  0.648$\pm$0.084 \\
\hline
 PAL\,15   & ${\it r}_{\rm C}$ sample     & 0.074$\pm$0.014  &  0.060$\pm$0.010 & 0.043$\pm$0.007 &  0.148$\pm$0.028 \\
           & ${\it r}_{\rm C-HM}$ sample     & ---  &  --- & --- & --- \\
           & ${\it r}_{\rm oHM}$ sample   & 0.081$\pm$0.013  &  0.075$\pm$0.010 &  0.055$\pm$0.006 &  0.162$\pm$0.026 \\
${\it R}_{\rm min}$=0.00 & WFC field                  & 0.074$\pm$0.010  &  0.066$\pm$0.007 &  0.049$\pm$0.004 &  0.148$\pm$0.020 \\
\hline
PYXIS      & ${\it r}_{\rm C}$ sample     & 0.096$\pm$0.014  &  0.085$\pm$0.011 & 0.074$\pm$0.008 &  0.192$\pm$0.028 \\
           & ${\it r}_{\rm C-HM}$ sample     & ---  &  --- & --- & --- \\
           & ${\it r}_{\rm oHM}$ sample   & 0.093$\pm$0.021  &  0.079$\pm$0.016 &  0.069$\pm$0.012 &  0.186$\pm$0.042 \\
${\it R}_{\rm min}$=0.00 & WFC field                  & 0.094$\pm$0.012  &  0.080$\pm$0.009 &  0.068$\pm$0.006 &  0.188$\pm$0.024 \\
\hline
RUP\,106   & ${\it r}_{\rm C}$ sample    & 0.156$\pm$0.010   &  0.130$\pm$0.008 & 0.098$\pm$0.006 &  0.312$\pm$0.020 \\
           & ${\it r}_{\rm C-HM}$ sample  & 0.152$\pm$0.039  &  0.120$\pm$0.030 & 0.087$\pm$0.023 &  0.304$\pm$0.078 \\
           & ${\it r}_{\rm oHM}$ sample   & 0.137$\pm$0.009  &  0.113$\pm$0.007 &  0.076$\pm$0.005 &  0.274$\pm$0.018 \\
${\it R}_{\rm min}$=0.00 & WFC field                  & 0.146$\pm$0.007  &  0.121$\pm$0.005 &  0.086$\pm$0.004 &  0.292$\pm$0.014 \\
\hline
\hline
\noalign{\smallskip}
\end{tabular}
\end{table*}

\subsection{Radial distribution}
\begin{centering}
\begin{figure*}
 \includegraphics[width=15.5cm]{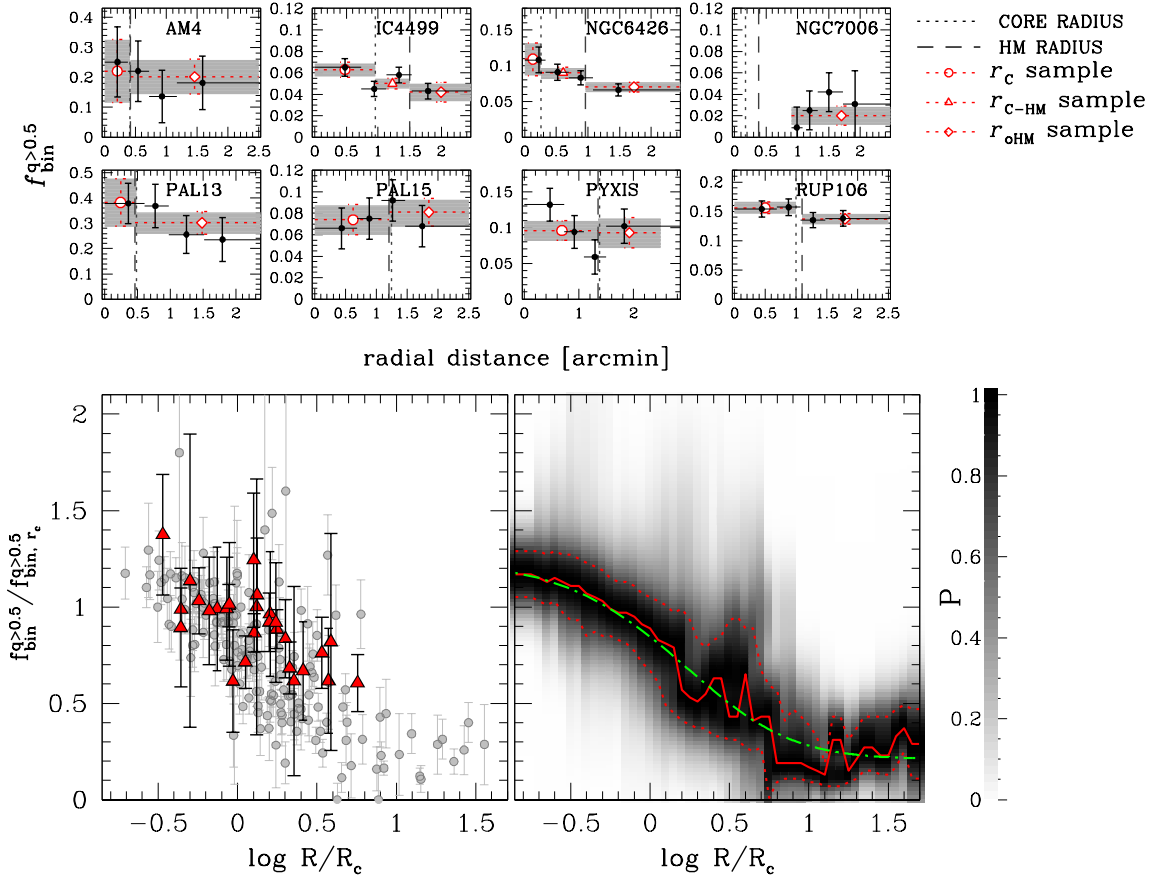}
 \caption{\textit{Upper panels:} Fraction of binaries with mass ratio q$>$0.5 as a function of the radial distance from the cluster center. The dotted and dashed vertical black lines mark the core and the half-mass radius. The fraction of binaries in the $r_{\rm C}$, $r_{\rm C-HM}$, and $r_{\rm oHM}$ sample are plotted with red symbols, while the corresponding errors are indicated by red error bars and shadowed areas. The radial interval corresponding to each red symbol is represented with a red dashed segment. Black dots indicate the fraction of binaries in four radial intervals and the radial coverage of each point is indicated by black horizontal segments. \textit{Lower-left panel:} Binary fraction with q$>$0.5, in units of core binaries as a function of the radial distance in units of core radii. Red triangles and grey dots indicate the clusters studied in this paper and in MPB+12, respectively. \textit{Lower-right panel:} The gray levels mark the normalised probability distribution in the same plane as in the lower-left panel. The red continuous line connect the points with probability, P=1.00, while the dotted lines represent the one-sigma confidence intervals. The green dash-dotted line is the best fitting function of the points with  probability equal to one. See text for details.}
 \label{fig:RD}
\end{figure*}
\end{centering}

 To investigate the radial distribution of binaries, we follow the same recipe as described in Sect.~5.5 of MPB+12. Briefly, we have divided the field of view in four concentric annuli, each containing the same number of stars in the CMD region $A$, and determined the fraction of binaries in each annulus. 
 Results are illustrated in Fig.~\ref{fig:RD}.
 Upper panels show  $f_{\rm bin}^{\rm q>0.5}$ for the eight GCs studied in this paper. In six clusters the fraction of binaries is maximum in the innermost cluster regions and decreases at larger radial distances, while in the case of Pal\,15 we observe a flat distribution. Unfortunately, stellar crowding prevented us from estimating the binary fraction in the central region of NGC\,7006.

 In the lower-left panel of Fig.~\ref{fig:RD}, we compare the fraction of binaries, normalised to the core binary fraction, for the clusters studied by MPB+12 (gray dots) and for those of this paper (red triangles). As already noticed by MPB+12, it seems that binaries in GCs follow a common radial trend.  

To further investigate this issue  we have considered a series of log(R/R$_{\rm c}$) intervals with a width of 0.25. These intervals are defined over a grid of points separated by 
 log(R/R$_{\rm c}$)=0.05 dex.   
For simplicity, we indicate with $f_{\rm n,i}$, where i=1,..,N, the normalised fraction of binaries within each radial bin ($f_{\rm bin}^{\rm q>0.5}/f_{\rm bin, r_{C}}^{\rm q>0.5}$). Then we have defined a vector, $f$, whose components range from $f$=0.00 to $f$=2.20 in steps of 0.01. For each component, j, we have calculated:\\
$p_{\rm j}=\sum_{\rm i=1}^{\rm N}  \exp{\frac{-(f_{\rm n,i}-f_{\rm j})^{2}}{2 \sigma_{\rm i}^{2}}}$ \\
 where $\sigma_{\rm i}$ is the error associated to $f_{\rm n,i}$. 
 The normalised probability for each radial bin, $P_{\rm j}$, is estimated by dividing $p_{\rm j}$ by the maximum value of $p_{\rm j}$ in that bin. 

Results are illustrated in the lower-right panel of Fig.~\ref{fig:RD} where the gray levels indicate the normalised probability in the $f_{\rm bin}^{\rm q>0.5}/f_{\rm bin, r_{\rm C}}^{\rm q>0.5}$ vs.\,log(R/R$_{\rm c}$) plane. We have connected with  continuous and dotted red lines the points with $P=1.00$ and $P=0.68$, respectively. 
The continuous line reaches its maximum within the core, and drops down by a factor of two at log(R/R$_{\rm c}$)$\sim$0.3. The maximum probability reaches its minimum at radial distance of $\sim$5 core radii, and it is almost constant at larger radii. This behaviour supports the idea that, when scaled as in Fig.~\ref{fig:RD}, there is a common trend in binary fraction with the core radius in all the GCs.  
 The points with $P=1.00$ are well reproduced by the function:
\begin{equation}
  f(R^{*})= \frac{a1}{\big{(}1+\frac{R^{*}}{a2}\big{)}^{2}}+a3
\end{equation}
where $f=f_{\rm bin}^{\rm q>0.5}/f_{\rm bin, r_{\rm C}}^{\rm q>0.5}$, $R^{*}=R/R_{\rm C}$, and the best-fitting values for the constants are $a1$=1.05, $a2$=3.5, and $a3$=0.2.  
 
 As mentioned above, in order to derive the results illustrated in Fig.~\ref{fig:RD} we have divided the field of view in four regions with the same number of stars. To test whether our conclusion depends on the adopted binning or not,  we have divided the field of view of each cluster in N concentric annuli, such that the minimum and maximum radius of each of them corresponds to (i$-$1)$ \times r_{\rm C}$ and i$\times r_{\rm C}$, respectively, where $r_{\rm C}$ is the core radius of the cluster and i=1,2,...,N. 

\begin{centering}
\begin{figure*}
 \includegraphics[width=15.5cm]{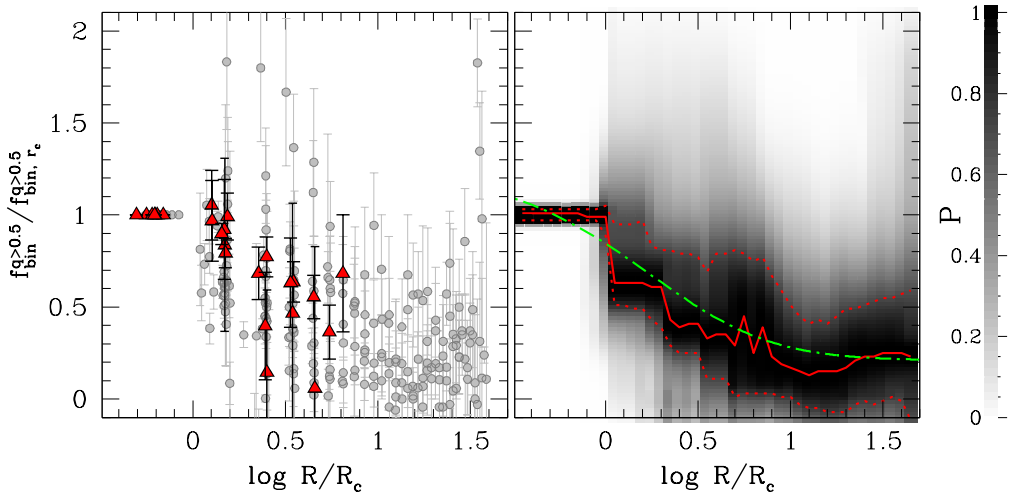}
 \caption{ As in the lower panels of Fig.~\ref{fig:RD} but in this case we have adopted annuli, whose radii are multiples of the cluster core radius. The green dashed-dotted line in the right panel is the best-fitting function derived from Fig.~\ref{fig:RD}. See text for details.}
 \label{fig:RDbinCORE}
\end{figure*}
\end{centering}

 We have thus estimated the fraction of binaries in each annulus for both the GCs studied in this paper and those from MPB+12 and illustrate the results in Fig.~\ref{fig:RDbinCORE}. The comparison of the fraction of binaries, normalised to the core binary fraction is illustrated in the left panel of Fig.~\ref{fig:RDbinCORE} and confirms the impression that, on average, GCs follows a common trend.
 The normalised probability in the $f_{\rm bin}^{\rm q>0.5}/f_{\rm bin, r_{\rm C}}^{\rm q>0.5}$ vs.\,log(R/R$_{\rm c}$) plane is plotted in the right panel of Fig.~\ref{fig:RDbinCORE} and has been derived as described above. Similarly the points with $P=1.00$ and $P=0.68$ have been connected with continuous and dotted red lines, respectively.

The green dashed-dotted line superimposed on this diagram is the best fitting function derived from Fig.~\ref{fig:RD}. The fact that this relation closely reproduces the average observed trend demonstrates that results are not significantly affected by the adopted binning.

\subsection{Mass-ratio distribution}
In their study of the mass-ratio distribution of the binary populations, MPB+12 has divided the region B of the CMD into five mass-ratio bins each covering a given mass-ratio interval ($\Delta$q$_{\rm i}$, i=1,2,...,5). The size of the five regions has been determined with the criterion that each of them cover almost the same area in the CMD.  Specifically, we have used q$_{\rm min, i}$=0.5, 0.61, 0.69, 0.76, 0.83 and q$_{\rm max, i}$=0.61, 0.69, 0.76, 0.83, 1.0 as minimum and maximum values of the mass ratio, respectively, for the five analysed regions. (see Sect.~5.1 of MPB+12 for  further details).   
 MPB+12 has calculated the fraction of binaries over the entire field of view, and have the normalised fraction of binaries $\nu_{\rm bin, i}=f_{\rm bin, i}/\Delta {\rm q}_{\rm i}$. This normalisation allows to account for the different mass-ratio intervals spanned by each region and properly investigate the mass-ratio distribution.
 Moreover, to increase the statistical sample, they have divided the region B into two subregions with 0.5$<$q$<$0.7 and 0.7$<$q$<$1.0 and calculated $\nu_{\rm bin}$ in each of them.

\begin{centering}
\begin{figure*}
 \includegraphics[width=15cm]{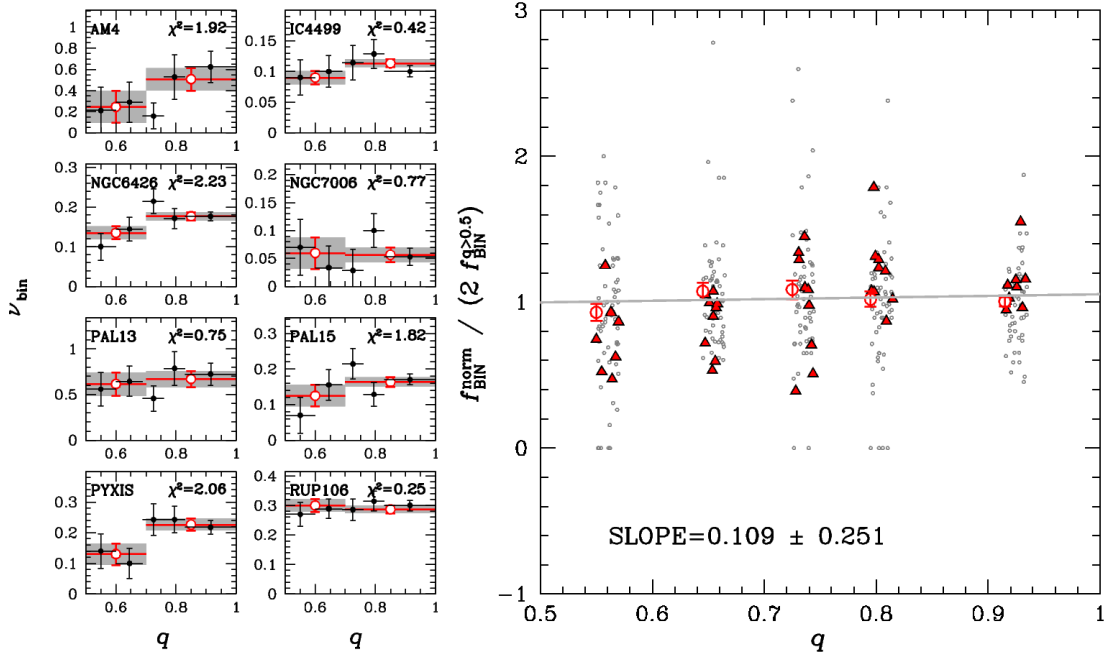}
 \caption{\textit{Left panels:} Mass ratio distribution of binaries in the WFC/ACS or WFC3/UVIS field for the GCs studied in this paper. The normalised fraction of binaries in five mass-ratio intervals are represented with black circles, while red-open circles indicate the values of $\nu_{\rm bin}$ measured in two bins with 0.5$<$q$\leq$0.7 and 0.7$<$q$\leq$1.0. The mass-ratio interval corresponding to each point is plotted with horizontal segments, observational errors are represented with vertical bars and shadowed area.
 \textit{Right panel:} Normalised fraction of binaries with q$>$0.5, $\nu_{\rm bin}$, as a function of the mass ratio, q. The fraction of binaries have been calculated in five intervals of mass ratio, and in order to compare binaries in different clusters we have divided $\nu_{\rm bin}$ by two times $f_{\rm BIN}^{\rm q>0.5}$. Gray points and red triangles indicate the values obtained by MPB+12 and in this paper, respectively. The mean normalised binary fraction in each mass-ratio bin are plotted with red circles. The gray line is the best-fit line, whose slope is indicated in the inset.  
 }
 \label{fig:QD}
\end{figure*}
\end{centering}

In this section, we extend the analysis by MPB+12 to the outer-halo GCs studied in this paper and plotted $\nu_{\rm bin, i}$ vs.\,q for each cluster in the left panels of Fig.~\ref{fig:QD}. We also compared observations with a flat distribution and calculated for each cluster the reduced $\chi^{2}$. We find that all GCs exhibit a nearly flat mass-ratio distribution. In the cases of Pyxis, AM\,4, NGC\,6426, and Pal\,15 there is a larger fraction of binaries with q$>$0.7, although this result is only significant at the level of $\sim$1-2 $\sigma$.

 In order to compare results from different clusters we have normalised $\nu_{\rm bin, i}$ by two times the fraction of binaries with q$>$0.5 estimated in the entire field of view. We have used gray points to plot points from the 59 GCs studied by MPB+12, while red triangles indicate the eight GCs of this paper. For clarity, points have been scattered around the corresponding q value. 
 For each value of q$_{\rm i}$, we have determined the mean normalised $\nu_{\rm bin, i}$ and plotted it with red circles. The red error bars are obtained by dividing the r.m.s. of the normalised $\nu_{\rm bin, i}$ by the square-root of $N-1$ where, $N$ is the number of analysed clusters. These mean points have been fitted with a straight line by means of least-squares and the best fitting line is represented with gray color code in Fig.~\ref{fig:QD}. Its slope is quoted in the figure and is consistent with zero, thus confirming the conclusion by MPB+12 that, on average the mass-ratio distribution of binaries in GCs is flat.

\subsection{The binary fraction as a function of  primary-star luminosity and mass}
In order to investigate the dependence of the binary fraction from the luminosity of the primary star we have calculated the fraction of binaries with q$>$0.5 over the entire field of view in three F814W magnitude intervals, in close analogy with what was done in MPB+12 (see their Sect.~5.4). Specifically, we have used three luminosity bin including all the single stars and the binaries with a primary star [0.75,1.75], [1.75,2.75] and [2.75,3.75] F814W magnitude fainter than the MSTO, respectively. We will indicate as $f_{\rm bin, b}^{\rm q>0.5}$,  $f_{\rm bin, i}^{\rm q>0.5}$,  $f_{\rm bin, f}^{\rm q>0.5}$ the fraction of binaries with q$>$0.5 measured in the bright, intermediate, and faint luminosity interval, respectively.

\begin{centering}
\begin{figure*}
 \includegraphics[width=15cm]{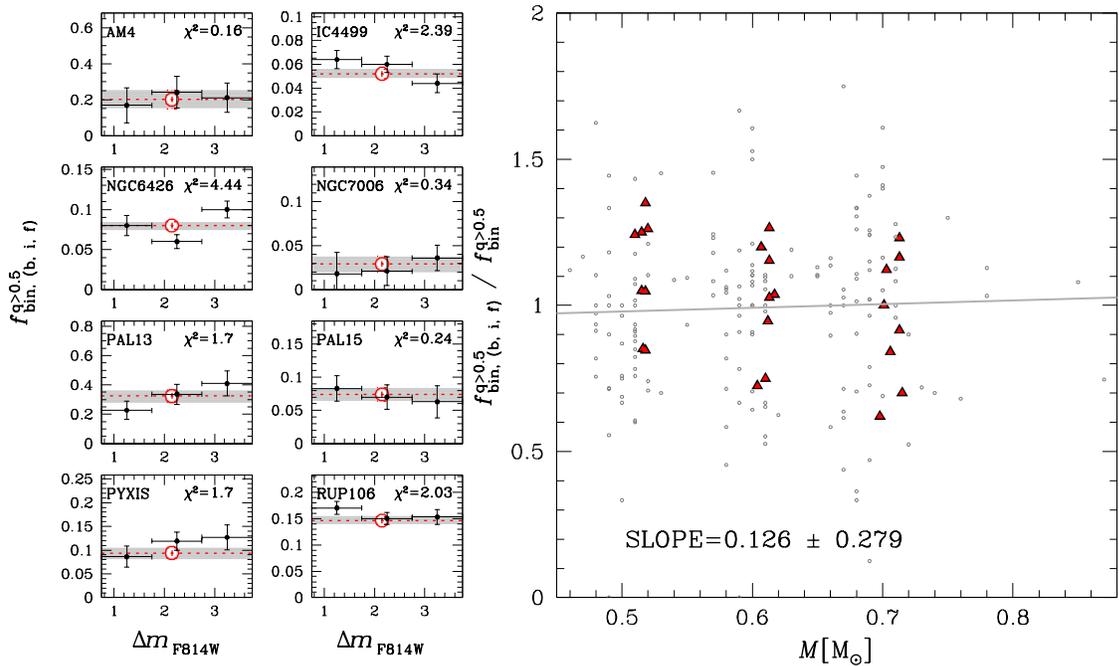}
 \caption{\textit{Left panels:} Black points show the fraction of binaries with q$>$0.5, measured in three magnitude intervals in the ACS or UVIS field of view as a function of the F814W magnitude distance from the MS TO. The fraction of binaries with q$>$0.5 in the interval between 0.75 and 3.75 F814W below the MS TO is represented with red circles.  Horizontal segments indicate the magnitude coverage corresponding to each point. \textit{Right panel:} Fraction of binaries with q$>$0.5 measured in three magnitude intervals, normalised to $f_{\rm bin}^{\rm q>0.5}$ as a function of the mass of the primary component. Symbols are like in the left panel of Fig.~\ref{fig:QD}.}
 \label{fig:MD}
\end{figure*}
\end{centering}

Results are illustrated in Fig.~\ref{fig:MD}. The black dots in the left panels show $f_{\rm bin, b,i,f}^{\rm q>0.5}$ as a function of the F814W magnitude difference from the MSTO ($\Delta m_{\rm F814W}$). The red circles indicate the fraction of binaries with q$>$0.5 in the entire [0.75,3.75] magnitude interval. Horizontal lines mark the $\Delta m_{\rm F814W}$ interval corresponding to each points, while the vertical lines and the shadowed areas represent error bars.

To compare results from different clusters and investigate the relation between the fraction of binaries and the mass of the primary component, in the right panel of Fig.~\ref{fig:MD} we have divided $f_{\rm bin, b,i,f}^{\rm q>0.5}$  by the fraction of binaries with q$>$0.5 in the entire field  and plotted $f_{\rm bin, b,i,f}^{\rm q>0.5}$ against the average mass of the primary-component stars of the binaries in each magnitude interval. Stellar masses have been derived by using the mass-luminosity relation provided by the adopted isochrones from Dotter et al.\,(2008). Gray points and red triangles indicate results from this paper and from MPB+12. 
 The gray line is the best-fitting straight line, whose slope, indicated in the inset support the conclusion by MPB+12 that the binary fraction does not depend on the mass of the primary component.

\section{Relation between binary fraction and globular-cluster parameters}\label{sec:relations}
In this Section we investigate correlation between the binary fraction and the physical and some morphological parameters of the host GCs listed in Sect.~\ref{subs:parameters}, in close analogy with what done by MPB+12. Results are provided in Table~\ref{tab:risultati} where we indicate for each parameter the Spearman's rank correlation coefficient, $r$. 
We used bootstrapping statistics to estimate uncertainties  in $r$. To do this we generated 1,000 resamples of the observed dataset, of equal size, and for each resample (i), (which is generated by random sampling with replacement from the original dataset) we estimated r$_{\rm i}$.  We considered the dispersion of the r$_{\rm i}$ measurements ($\sigma_{\rm r}$) as indicative of robustness of r and list the number of  included GCs ($N$).
Since the definition of core radius is not reliable for post-core-collapse (PCC) clusters  (Trager et al.\,1995) we have excluded these GCs from the calculation of $r$.

\begin{centering}
\begin{figure*}
 \includegraphics[width=12cm]{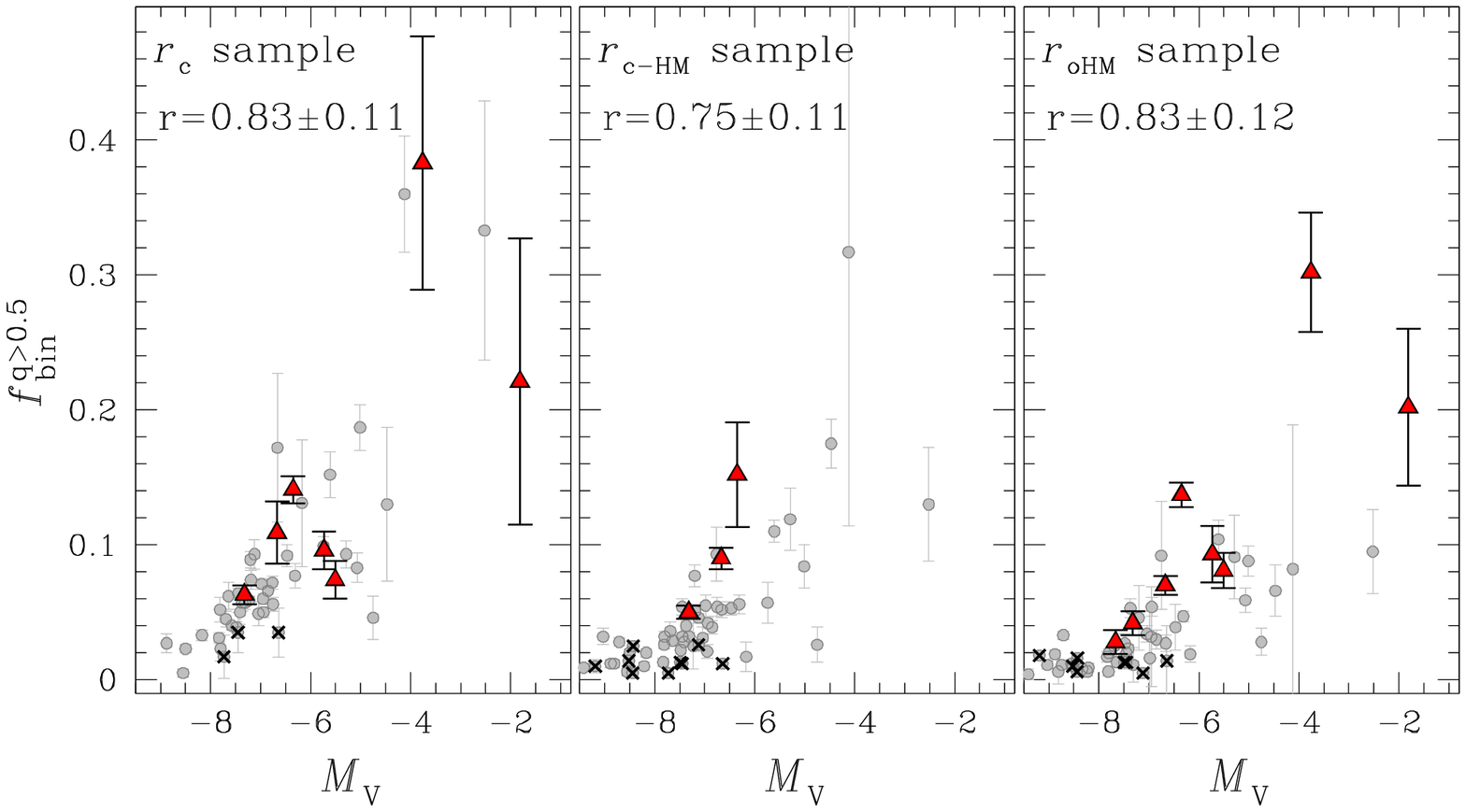}
 \caption{ Fraction of binaries with q$>$0.5 as a function of the absolute magnitude for stars in the $r_{\rm c}$ (left), $r_{\rm c-HM}$ (middle), and $r_{\rm oHM}$ sample (right). Gray dots and red triangles indicate clusters studied by MPB+12 and in this paper, respectively, while PCC clusters are marked with black crosses. The Spearman correlation coefficient, $r$, is indicated in each panel. }
 \label{fig:fMv}
\end{figure*}
\end{centering}

 Figure~\ref{fig:fMv} shows a significant correlation between the fraction of binaries and the absolute magnitude of the host GC, which is present in each $r_{\rm c}$, $r_{\rm c-HM}$, and $r_{\rm oHM}$ sample. This relation is clearly visible from the clusters studied by MPB+12 (gray dots) and the eight GCs studied in this paper (red triangles) follow a similar trend in the $f_{\rm bin}$ vs.\,$M_{\rm V}$ plane. In particular the two GCs with very-low luminosities, namely AM\,4 and Pal\,13 host a large fraction of binaries with mass ratio larger than 0.5 ($f_{\rm bin}^{\rm q>0.5}>$0.2). Noticeable, the presence of a large binary fraction in Pal\,13 confirms previous findings by Clark\, Sandquist \& Bolte\,2004 who analysed photometry from Keck\,II telescope of this cluster and concluded that Pal\,13 host more than 30$\pm$4\% of binaries. 

In contrast, PCC clusters, which are represented with black crosses, seem to exhibit a flat distribution in the middle and left panel of Fig.~\ref{fig:fMv}. 
However, we admit that this conclusion is currently based only on a small number of PCC GCs (nine in total) and a small interval of absolute luminosity.
  Moreover, there is a significant scatter for clusters with the the same absolute luminosity, which suggest that apart from the GC luminosity (a proxy for the GC mass) at least one more parameter is needed to explain the content of binaries in GCs. 

\begin{centering}
\begin{figure*}
 \includegraphics[width=9cm]{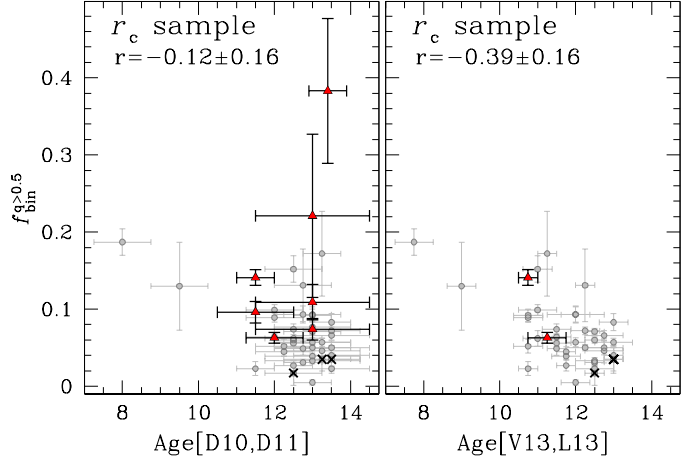}
 \caption{Fraction of binaries with q$>$0.5 in the core as a function of the ages of the host cluster. The ages used in the left- and right panel are from Dotter et al.\,(2010, 2011) and from VandenBerg et al.\,(2013) and Leaman et al.\,(2013), respectively.}
 \label{fig:ages}
\end{figure*}
\end{centering}

The correlation between the binary fraction and the age of the host cluster has been widely debated in recent literature.
 Ji \& Bregman\,(2015) have estimated the fraction of binaries in 35 Galactic GCs from the same dataset by Sarajedini et al.\,(2007) and Anderson et al.\,(2008) previously analysed by MPB+12. From the analysis of the binary fraction in the core of 24 GCs, they concluded that the fraction of binaries in the core decreases with the cluster age as previously suggested by Sollima et al.\,(2007) on the basis of their analysis of 13 low-density GCs.
 These results are not in agreement with the conclusion by MPB+12 who did not find any significant correlation between the fraction of binaries and the cluster ages by Mar{\'{\i}}n-Franch et al.\,(2009) and De Angeli et al.\,(2005).
We have plotted, in Fig.~\ref{fig:ages}, the fraction of binaries with q$>$0.5 in the core as a function of the cluster ages derived by Dotter et al.\,(2010, 2011 left panel) and by Vandenberg et al.\,(2013) and Leaman et al.\,(2013, right panel)\footnote{ Unfortunately, the sample of clusters studied by  Mar{\'{\i}}n-Franch et al.\,(2009) and De Angeli et al.\,(2005) do not include any of the eight GCs studied in this paper.}. 
Our results do not confirm the finding by Ji \& Bregman, indeed,  as quoted in each panel, the Spearman's rank correlation coefficient is $r \sim -$0.1 for ages by Dotter et al.\,(2011) and $r \sim -$0.4 when ages from  Vandenberg, Leaman and collaborators are used. Such difference, is in part due to the fact that the latter age compilation includes less clusters than those analysed by Dotter et al.\,(2010, 2011).  Similarly, the disagreement between our conclusions and those by Ji \& Bregman\,(2015) could be due to the fact that some old clusters like E\,3, AM\,4 or Pal\,13 with large binary fractions are not included in the sample analysed by Ji \& Bregman. 

Although binaries have been considered as a candidate second-parameter to explain the HB morphology in GCs (e.g.\,Napiwotzki et al.\,2004; Lei et al.\,2014), MPB+12 did not find any significant relation between the fraction of binaries and the HB parameters that they have analysed. These include the HB morphology index (HBR, from Mackey \& van den Bergh 2005), the median color difference between the HB and the RGB, $\Delta$(V$-$I), from Dotter et al.\,2010), and the temperature of the hottest HB star ($T_{\rm eff, HB}$, from Recio-Blanco et al.\,2006). 
 When we extend the analysis to the GCs studied in this paper, we confirm the lack of a significant correlation between the fraction of binaries and HBR.
 Unfortunately, there are no estimates of $T_{\rm eff, HB}$ and  $\Delta$(V$-$I) available the eight clusters analysed in this work.  
In addition to what has been done in MPB+12, we investigate the correlation between the binary fraction and the color distance between the reddest part of the HB and the RGB (L$_{1}$) and the color extension of the HB (L$_{2}$, from Milone et al.\,2014). There is a very marginal $f_{\rm bin}$ vs.\,L$_{1}$ anticorrelation (r=$-$0.41) and some hint of anticorrelation between $f_{\rm bin}$ and  L$_{2}$ (r=$-$0.58), although this latter correlation is expected given that L$_2$ correlates with the cluster absolute magnitude (Milone et al.\,2014). 

Finally, we note that there is no correlation between $f_{\rm bin}$ and the following parameters of the host cluster: [Fe/H], c, $\rho_{0}$, log($\tau_{c}$), log($\tau_{hm}$), e, R$_{\rm GC}$, R, S$_{\rm RR Lyrae}$. The fraction of binaries anti-correlates with the central velocity dispersion, and there is a mild correlation with $\mu_{\rm V}$. These are a consequence of the fact that the later two quantities correlates with M$_{\rm V}$.

\begin{centering}
\begin{figure*}
 \includegraphics[width=15cm]{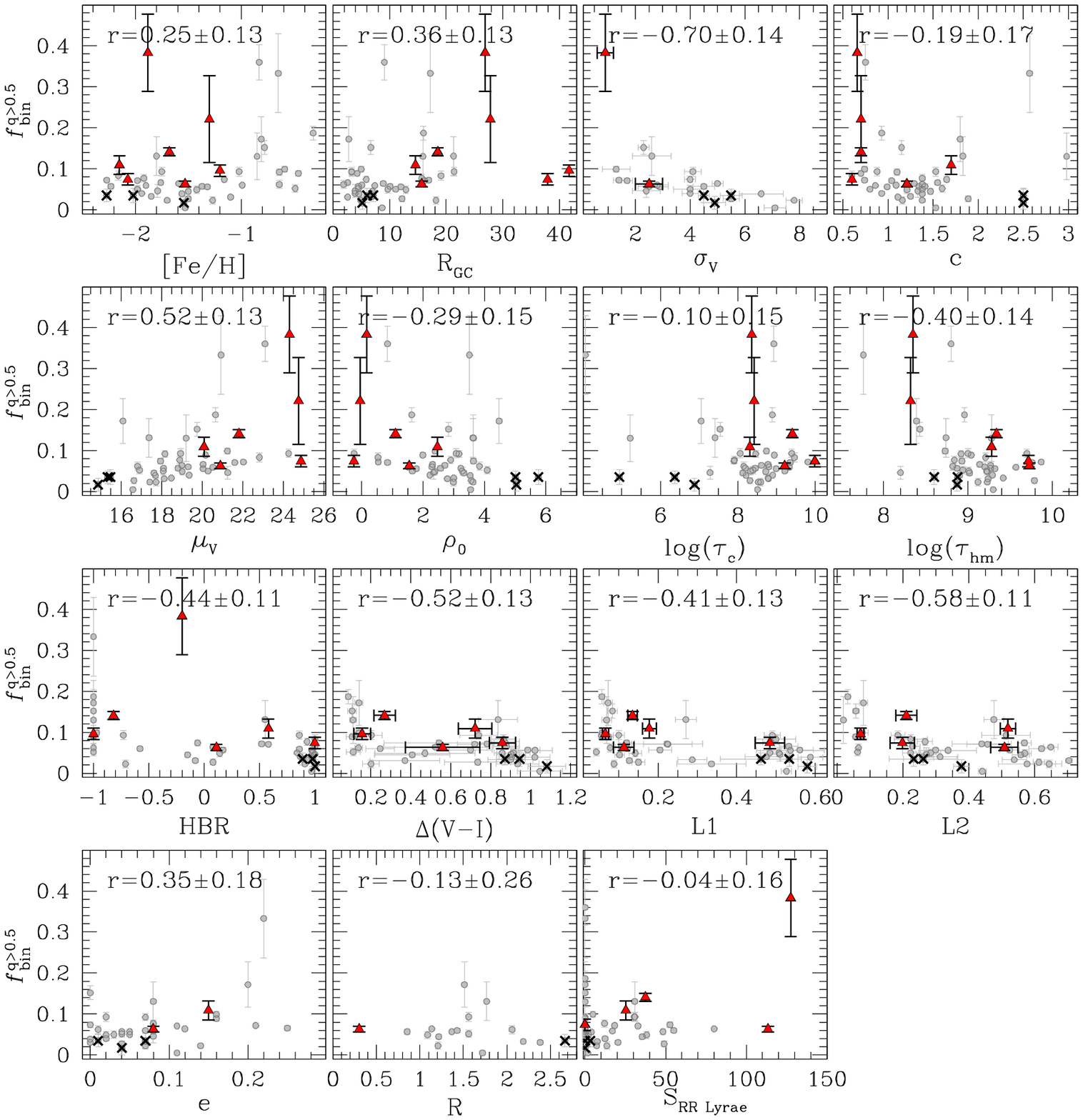}
 \caption{Fraction of binaries with q$>$0.5 in the core as a function of some parameters of their host GCs. From the top-left: metallicity, age, central velocity dispersion, King-model central concentration, central surface brightness, central luminosity density, logarithm of core relaxation time, logarithm of median relaxation time, ellipticity, distance from the Galactic center, R parameter, frequency of RR\,Lyrae, HB ratio, $m_{\rm F606W}-m_{\rm F814W}$ color distance from the reddest part of the HB and the RGB, $m_{\rm F606W}-m_{\rm F814W}$ color extension of the HB. Symbols are the same as in Fig.~\ref{fig:fMv}.}
 \label{fig:REL}
\end{figure*}
\end{centering}

\begin{table*}
\label{tab:risultati}
\centering
\caption{
Spearman's rank correlation coefficients indicating the statistical dependence between the fraction of binaries with q$>$0.5 and several parameters of the host GCs for ${\it r}_{\rm c}$, ${\it r}_{\rm c-HM}$, and ${\it r}_{\rm oHM}$ sample. The uncertainty of $r$ measurements, and the numbers of analysed GCs (N$_{\rm CL}$) are also listed.}
\begin{tabular}{cccccccc}
\hline
\hline
\noalign{\smallskip}
\multicolumn{1}{c}{Parameter} &  
\multicolumn{1}{c}{${\it r}_{\rm c}$} & 
\multicolumn{1}{c}{} & 
\multicolumn{1}{c}{${\it r}_{\rm c-HM}$} &    
\multicolumn{1}{c}{} & 
\multicolumn{1}{c}{${\it r}_{\rm oHM}$} & 
\multicolumn{1}{c}{}      \\
\hline
  &   $r$   & N$_{\rm CL}$     & $r$ & N$_{\rm CL}$  & $r$ & N$_{\rm CL}$   \\
\hline
    $M_{\rm V}$    &    0.83$\pm$0.11     & 46 &     0.75$\pm$0.11 & 46 & 0.83$\pm$0.11  & 48 \\
     Age[D10,D11] & $-$0.12$\pm$0.16     & 43 &  $-$0.25$\pm$0.15 & 43 & $-$0.04$\pm$0.16  & 46 \\
     Age[V13,L13] & $-$0.39$\pm$0.16     & 37 &  $-$0.45$\pm$0.12 & 41 & $-$0.22$\pm$0.17  & 37 \\
    $\rm [Fe/H]$  &    0.25$\pm$0.13     & 46 &     0.17$\pm$0.15 & 46 & $-$0.08$\pm$0.15  & 48 \\
     R$_{\rm GC}$   &   0.36$\pm$0.13     & 46 &     0.31$\pm$0.13 & 46 &    0.52$\pm$0.11  & 48 \\
$\sigma_{\rm V}$   & $-$0.70$\pm$0.14   & 21 &  $-$0.68$\pm$0.12 & 25 & $-$0.78$\pm$0.08    & 24 \\
     $c$          & $-$0.19$\pm$0.17   & 45 &  $-$0.37$\pm$0.16 & 46 & $-$0.57$\pm$0.12    & 47 \\
    $\mu_{\rm V}$    &  0.52$\pm$0.13   & 45 &     0.71$\pm$0.09 & 46 &    0.85$\pm$0.05  & 47 \\
    $\rho_{0}$      & $-$0.29$\pm$0.15   & 45 &  $-$0.50$\pm$0.13 & 46 & $-$0.77$\pm$0.07  & 47 \\
log($\tau_{\rm c}$)  & $-$0.10$\pm$0.15   & 45 &     0.19$\pm$0.17 & 46 &    0.36$\pm$0.15  & 47 \\
log($\tau_{\rm hm}$) & $-$0.40$\pm$0.14   & 45 &  $-$0.16$\pm$0.16 & 46 &    0.01$\pm$0.17  & 47 \\
      HBR           & $-$0.44$\pm$0.11   & 44 &  $-$0.34$\pm$0.14 & 45 & $-$0.18$\pm$0.16 & 44 \\
$\Delta$(V$-$I)     & $-$0.52$\pm$0.13   & 41 &  $-$0.40$\pm$0.14 & 43 & $-$0.18$\pm$0.17 & 44 \\
      L1            & $-$0.41$\pm$0.13   & 41 &  $-$0.26$\pm$0.14 & 43 &    0.06$\pm$0.17 & 44 \\
      L2            & $-$0.58$\pm$0.11   & 41 &  $-$0.45$\pm$0.13 & 43 & $-$0.57$\pm$0.12 & 44 \\
      $e$           &    0.35$\pm$0.18   & 34 &     0.25$\pm$0.18 & 39 &    0.17$\pm$0.17  & 36 \\
      $R$           & $-$0.13$\pm$0.26   & 16 &  $-$0.54$\pm$0.15 & 21 & $-$0.36$\pm$0.25  & 21 \\
S$_{\rm RR~Lyrae}$    & $-$0.04$\pm$0.16   & 43 &  $-$0.13$\pm$0.14 & 46 &    0.01$\pm$0.15  & 45 \\
\hline
\hline
\noalign{\smallskip}
\end{tabular}
\end{table*}

\section{Summary}
We have used archive images from UVIS/WFC3 and WFC/ACS to study the population of MS-MS binaries in eight GCs with distance from the Galactic center larger than $\sim$15 kpc. We have measured the fraction of binaries with high mass-ratio q$>$0.5 and extrapolated the fraction of binaries with q$<$0.5 in different regions. Specifically we have calculated the fraction of binaries in the entire field of view, within the core, in the region between the core and the half-mass radius, and outside the half-mass radius. We studied the radial and mass-ratio distribution of binaries and the distribution of binary systems as a function of the luminosity and the mass of the primary star. Moreover, we have investigated monovariate relations between the fraction of binaries and the main parameters of the host GC. This work thus extends the analysis by MPB+12 who have investigated binaries in 59 Galactic GCs by using the same method.

We found that the binary fraction varies from one cluster to another and changes, within the same cluster, when different regions at different radial distances are analysed. As an example, the fraction of binaries with q$>$0.5 ranges from $\sim$0.38, in the core of Pal\,13 to $\sim$0.03 outside the half-mass radius of NGC\,7006.
 
Our investigation of the radial distribution supports the conclusion by MPB+12 that the binary fraction in GCs, when normalised to the core binaries follow a common radial trend. 
Specifically, by combining results from this paper and from MPB+12, we find that the binary fraction is maximum in the innermost cluster regions, drops by approximately a factor of two, at a distance of two core radii and approach its minimum  of $\sim$20\% at larger radial distances of about five core radii.
We find that, when results from all the clusters are studied together, both the average mass-ratio distribution and the average distribution of binaries with respect to the  mass of primary stars are flat.

 Our sample includes two very-low luminosity clusters, AM\,4 and Pal\,13, with $M_{\rm V}>-4$ which exhibit both a very high binary fraction ($f_{\rm bin}^{\rm q>0.5}> \sim$0.2). The fraction of binaries decreases in GCs with brighter absolute luminosities, thus confirming the correlation between $f_{\rm bin}$ and $M_{\rm V}$ from MPB+12. 

 In this context, it is worth mentioning the work by Fadely et al.\,(2011) who studied the ultra faint ($M_{\rm V}=0.0 \pm 0.8$) star cluster Segue\,3 in the outer halo and detected only one spectroscopic binary over 32 analysed stars. 
 This results could suggest that Segue\,3 hosts a small binary fraction, in contrast with what observed for the low-luminosity clusters AM\,4, Pal\,13, and E\,3.  The so-called class of ultra-faint clusters with $M_{\rm V} >-1.0$  includes a small but increasing number of objects (e.g.\,Koposov et al.\,2007; Mu{\~n}oz et al.\,2012; Balbinot et al.\,2013; Kim \& Jerjen\,2015; Kim et al.\,2015)  but unfortunately their population of binary systems is still poorly known. 
 The determination of the binary fraction in these object is mandatory to firmly establish if the anticorrelation between the absolute luminosity and the binary fraction is still valid for $M_{\rm V} >-1.0$.

Interestingly, PCC clusters seem to not follow this relation, although the small number of available clusters prevent us from any strong conclusion. 
 As expected, the binary fraction also anticorrelates with the central velocity dispersion and and the color extension of the HB, L$_{2}$, and mildly correlates with $\mu_{\rm V}$, indeed these quantities are related with the absolute luminosity. 
We have investigated monovariate relations between the fraction of binaries and other parameters of the host GCs and did not find any significant correlation with  metallicity, age, central concentration, central luminosity density, core and median relaxation time, ellipticity, distance from the Galactic center, R parameter, frequency of RR\,Lyrae, HB ratio, $m_{\rm F606W}-m_{\rm F814W}$ color distance from the reddest part of the HB and the RGB, L$_{1}$.

 \section*{acknowledgments}
\small
We are grateful to the referee for useful suggestions that have improved the quality of this manuscript.
We thank Ryan Goldsbury for the values for the cluster centers used in the paper, and Jay Anderson who have provided the software for astrometry and photometry.
APM and HJ acknowledge support by the Australian Research Council through Discovery Early Career Researcher Award DE150101816 and Discovery Project grant DP150100862.

\bibliographystyle{aa}

\begin{thebibliography}{}
\bibitem[Anderson \& King(2006)]{2006acs..rept....1A} Anderson, J., \& King, I.~R.\ 2006, Instrument Science Report ACS 2006-01, 34 pages, 1 

\bibitem[Anderson et al.(2008)]{2008AJ....135.2055A} Anderson, J., Sarajedini, A., Bedin, L.~R., et al.\ 2008, \aj, 135, 2055 

\bibitem[Anderson \& Bedin(2010)]{2010PASP..122.1035A} Anderson, J., \& Bedin, L.~R.\ 2010, \pasp, 122, 1035 

\bibitem[Balbinot et al.(2013)]{2013ApJ...767..101B} Balbinot, E., Santiago, B.~X., da Costa, L., et al.\ 2013, \apj, 767, 101 

\bibitem[Bedin et al.(2005)]{2005MNRAS.357.1038B} Bedin, L.~R., Cassisi, S., Castelli, F., et al.\ 2005, \mnras, 357, 1038 

\bibitem[Bellazzini et al.(2002)]{2002AJ....123.1509B} Bellazzini, M., Fusi Pecci, F., Messineo, M., Monaco, L., \& Rood, R.~T.\ 2002, \aj, 123, 1509 

\bibitem[Bellini et al.(2011)]{2011PASP..123..622B} Bellini, A., Anderson, J., \& Bedin, L.~R.\ 2011, \pasp, 123, 622 

\bibitem[Bolte(1992)]{1992ApJS...82..145B} Bolte, M.\ 1992, \apjs, 82, 145 

\bibitem[Carretta et al.(2009)]{2009A&A...505..117C} Carretta, E., Bragaglia, A., Gratton, R.~G., et al.\ 2009, \aap, 505, 117 

\bibitem[Clark et al.(2004)]{2004AJ....128.3019C} Clark, L.~L., Sandquist, E.~L., \& Bolte, M.\ 2004, \aj, 128, 3019 

\bibitem[De Angeli et al.(2005)]{2005AJ....130..116D} De Angeli, F., Piotto, G., Cassisi, S., et al.\ 2005, \aj, 130, 116 

\bibitem[Dotter et al.(2008)]{2008ApJS..178...89D} Dotter, A., Chaboyer, B., Jevremovi{\'c}, D., et al.\ 2008, \apjs, 178, 89 

\bibitem[Dotter et al.(2010)]{2010ApJ...708..698D} Dotter, A., Sarajedini, A., Anderson, J., et al.\ 2010, \apj, 708, 698 

\bibitem[Dotter et al.(2011)]{2011ApJ...738...74D} Dotter, A., Sarajedini, A., \& Anderson, J.\ 2011, \apj, 738, 74 

\bibitem[Fadely et al.(2011)]{2011AJ....142...88F} Fadely, R., Willman, B., Geha, M., et al.\ 2011, \aj, 142, 88 

\bibitem[Girardi et al.(2005)]{2005A&A...436..895G} Girardi, L., Groenewegen, M.~A.~T., Hatziminaoglou, E., \& da Costa, L.\ 2005, \aap, 436, 895 

\bibitem[Goldsbury et al.(2010)]{2010AJ....140.1830G} Goldsbury, R., Richer, H.~B., Anderson, J., et al.\ 2010, \aj, 140, 1830 

\bibitem[Hamren et al.(2013)]{2013AJ....146..116H} Hamren, K.~M., Smith, G.~H., Guhathakurta, P., et al.\ 2013, \aj, 146, 116 

\bibitem[Ji \& Bregman(2013)]{2013ApJ...768..158J} Ji, J., \& Bregman, J.~N.\ 2013, \apj, 768, 158 

\bibitem[Ji \& Bregman(2015)]{2015arXiv150500016J} Ji, J., \& Bregman, J.~N.\ 2015, arXiv:1505.00016 

\bibitem[Kim \& Jerjen(2015)]{2015ApJ...799...73K} Kim, D., \& Jerjen, H.\ 2015, \apj, 799, 73 

\bibitem[Kim et al.(2015)]{2015ApJ...803...63K} Kim, D., Jerjen, H., Milone, A.~P., Mackey, D., \& Da Costa, G.~S.\ 2015, \apj, 803, 63 

\bibitem[King(1962)]{1962AJ.....67..471K} King, I.\ 1962, \aj, 67, 471 

\bibitem[Koposov et al.(2007)]{2007ApJ...669..337K} Koposov, S., de Jong, J.~T.~A., Belokurov, V., et al.\ 2007, \apj, 669, 337 

\bibitem[Kroupa(2002)]{2002Sci...295...82K} Kroupa, P.\ 2002, Science, 295, 82 

\bibitem[Leaman et al.(2013)]{2013MNRAS.436..122L} Leaman, R., VandenBerg, D.~A., \& Mendel, J.~T.\ 2013, \mnras, 436, 122 

\bibitem[Lei et al.(2014)]{2014PASJ...66...82L} Lei, Z., Chen, X., Kang, X., Zhang, F., \& Han, Z.\ 2014, \pasj, 66, 82 

\bibitem[Mar{\'{\i}}n-Franch et al.(2009)]{2009ApJ...694.1498M} Mar{\'{\i}}n-Franch, A., Aparicio, A., Piotto, G., et al.\ 2009, \apj, 694, 1498 

\bibitem[Milone et al.(2009)]{2009A&A...497..755M} Milone, A.~P., Bedin, L.~R., Piotto, G., \& Anderson, J.\ 2009, \aap, 497, 755 

\bibitem[Milone et al.(2012)]{2012A&A...537A..77M} Milone, A.~P., Piotto, G., Bedin, L.~R., et al.\ 2012, \aap, 537, A77 

\bibitem[Milone et al.(2014)]{2014ApJ...785...21M} Milone, A.~P., Marino, A.~F., Dotter, A., et al.\ 2014, \apj, 785, 21 

\bibitem[Mu{\~n}oz et al.(2012)]{2012ApJ...753L..15M} Mu{\~n}oz, R.~R., Geha, M., C{\^o}t{\'e}, P., et al.\ 2012, \apjl, 753, L15 

\bibitem[Napiwotzki et al.(2004)]{2004Ap&SS.291..321N} Napiwotzki, R., Karl, C.~A., Lisker, T., et al.\ 2004, \apss, 291, 321 

\bibitem[Recio-Blanco et al.(2006)]{2006A&A...452..875R} Recio-Blanco, A., Aparicio, A., Piotto, G., de Angeli, F., \& Djorgovski, S.~G.\ 2006, \aap, 452, 875 

\bibitem[Richer et al.(2004)]{2004AJ....127.2771R} Richer, H.~B., Fahlman, G.~G., Brewer, J., et al.\ 2004, \aj, 127, 2771 

\bibitem[Romani \& Weinberg(1991)]{1991ApJ...372..487R} Romani, R.~W., \& Weinberg, M.~D.\ 1991, \apj, 372, 487

\bibitem[Rubenstein \& Bailyn(1997)]{1997ApJ...474..701R} Rubenstein, E.~P., \& Bailyn, C.~D.\ 1997, \apj, 474, 701 

\bibitem[Salaris et al.(2004)]{2004A&A...420..911S} Salaris, M., Riello, M., Cassisi, S., \& Piotto, G.\ 2004, \aap, 420, 911 

\bibitem[Sarajedini et al.(2007)]{2007AJ....133.1658S} Sarajedini, A., Bedin, L.~R., Chaboyer, B., et al.\ 2007, \aj, 133, 1658 

\bibitem[Sollima et al.(2007)]{2007MNRAS.380..781S} Sollima, A., Beccari, G., Ferraro, F.~R., Fusi Pecci, F., \& Sarajedini, A.\ 2007, \mnras, 380, 781 

\bibitem[Trager et al.(1995)]{1995AJ....109..218T} Trager, S.~C., King, I.~R., \& Djorgovski, S.\ 1995, \aj, 109, 218 

\bibitem[VandenBerg et al.(2013)]{2013ApJ...775..134V} VandenBerg, D.~A., Brogaard, K., Leaman, R., \& Casagrande, L.\ 2013, \apj, 775, 134 

\bibitem[Zhao \& Bailyn(2005)]{2005AJ....129.1934Z} Zhao, B., \& Bailyn, C.~D.\ 2005, \aj, 129, 1934 

\end{thebibliography}

\end{document}